\documentclass[final,3p,times,twocolumn]{elsarticle}
\usepackage{epsfig}
\usepackage{amsmath}
\usepackage[normalem]{ulem}

\renewcommand\sout{\bgroup \color{blue} \ULdepth=-.5ex \ULset}
  \biboptions{sort&compress}
\usepackage{color}
\usepackage{latexsym}
\usepackage{amssymb}
\usepackage{dsfont}
\usepackage{multirow}
\usepackage{epsfig}

\newcommand{\bea}{\begin{eqnarray}}
\newcommand{\eea}{\end{eqnarray}}
\newcommand{\be}{\begin{equation}}
\newcommand{\ee}{\end{equation}}

\newlength\savedwidth

\begin{document}
\begin{frontmatter}

\author[pavia]{B.~Pasquini}
\author[bochum]{M.V.~Polyakov}
\author[mainz]{M.~Vanderhaeghen}

\address[pavia]{Dipartimento di Fisica,
Universit\`a degli Studi di Pavia, Pavia, Italy, \\and Istituto Nazionale di Fisica Nucleare,
Sezione di Pavia, Pavia, Italy}
\address[bochum]{Institute of Theoretical Physics II, Ruhr University Bochum, Bochum, D44780 Germany}
\address[mainz]{Institut f\"ur Kernphysik, Johannes Gutenberg-Universit\"at, Mainz, Germany, \\
and PRISMA Cluster of Excellence, Johannes Gutenberg-Universit\"at,  Mainz, Germany}
\title{Dispersive evaluation of the D-term form factor in deeply virtual Compton scattering}

\begin{abstract}
We present a dispersive representation of the D-term form factor for hard exclusive reactions, using unsubtracted $t$-channel dispersion relations.
The $t$-channel unitarity relation is saturated with the contribution of two-pion intermediate states,  using  the two-pion distributions amplitude for the $\gamma^*\gamma\rightarrow \pi\pi$ subprocess 
and reconstructing the $\pi\pi\rightarrow N\bar N$ subprocess from available information on  pion-nucleon partial-wave helicity amplitudes.
Results for the D-term form factor as function of $t$ as well as  at $t=0$ are discussed in comparison with available model predictions and phenomenological parametrizations.
\end{abstract}

\begin{keyword}
dispersion relations, deeply virtual Compton scattering, generalized parton distributions

\PACS 13.60.Hb  \sep ,11.55.Fv \sep 13.60.-r 

\end{keyword}
\end{frontmatter}

\section{Introduction}
The D term was originally introduced to complete the parametrization of the generalized parton distributions (GPDs) in hard exclusive reactions in terms of double distributions, and restore the polynomiality property of the singlet moments of unpolarized GPDs~\cite{Polyakov:1999gs}. This term turned out to be  a  crucial  contribution in the phenomenological description of deeply virtual Compton scattering (DVCS) observables, where different forms 
have been assumed with parameters tuned to DVCS data~\cite{Guidal:2013rya,Kumericki:2007sa}. On the theoretical side, the D-term is poorly known, and information are available only from a few models, such as the chiral quark soliton 
model~\cite{Schweitzer:2002nm, Goeke:2007fp, Goeke:2007fq, Goeke:2001tz, Wakamatsu:2007uc}, the Skyrme model~\cite{Cebulla:2007ei}, a Regge-improved diquark model~\cite{Mueller:2011bk}, as well as a first moment from lattice simulations~\cite{Gockeler:2003jfa,Hagler:2007xi}. Interesting studies have been also performed for the nucleon in nuclear matter~\cite{Kim:2012ts,Jung:2014jja}, for $Q$-ball systems~\cite{Mai:2012yc,Mai:2012cx} and within  different variants of chiral perturbation theory~\cite{Belitsky:2002jp,Diehl:2006ya,Ando:2006sk,Dorati:2007bk,Wang:2010hp,Moiseeva:2012zi,Wein:2014wma}.
    Recently, the D-term form factor acquired a new significance in the dispersive representation of  DVCS 
amplitudes~\cite{Kumericki:2007sa,Pasquini:2001yy,Drechsel:2002ar,Anikin:2007yh,Teryaev:2005uj,Radyushkin:2011dh,Diehl:2007jb,Polyakov:2008xm,Goldstein:2009ks}.
In particular, it was shown that the DVCS amplitudes satisfy subtracted dispersion relations (DRs) at fixed $t$  with the subtraction function defined by the D-term form factor~\cite{Anikin:2007yh}.
In the present Letter we set up dispersion relations in the $t$ channel  for this subtraction function.
The advantage of this dispersive representation is to provide a microscopic interpretation of the physical content of the D-term form factor in terms of $t$-channel exchanges with the appropriate quantum numbers.
The plan of the Letter is as follows. In Section~\ref{sect:2}, we review the derivation of the $s$-channel subtracted dispersion relations for the DVCS amplitudes.
In Section~\ref{sect:3}, we derive $t$-channel DRs for the D-term form factor. 
The unitarity relation  for the $t$-channel amplitudes is saturated with  two-pion intermediate states, using the two-pion distribution amplitude for the $\gamma^*\gamma\rightarrow \pi\pi$ subprocess 
and reconstructing the $\pi\pi\rightarrow N\bar N$ subprocess from available information on  pion-nucleon partial-wave helicity amplitudes.
We then discuss  the dispersive predictions for the D-term form factor in Section~\ref{sec:4}, and we conclude summarizing our results.

\section{Subtracted dispersion relations in the $s$-channel}
\label{sect:2}
We consider the DVCS 
process 
\begin{align}
\gamma^*(q) \, N ( p)  \rightarrow \gamma(q') \, N(p'),
\end{align}
where the variables in brackets denote the four-momenta of the participating particles.
The familiar Mandelstam variables are 
\begin{eqnarray}
s=(p+q)^2,\quad t=(q-q')^2,\quad u=(q-p')^2,
\end{eqnarray}
and are constrained by $s+u+t=2M^2_N-Q^2$, with $M_N$ the nucleon mass and $Q^2=-q^2$.
We will consider the Bjorken regime,  where the photon 
virtuality $Q^2$ and $s$ are large, and 
 $-t \ll s,Q^2$.
\newline
\noindent
To calculate the DVCS amplitude, one starts from its definition as a nucleon matrix element of the $T$-product of two electromagnetic currents:
\begin{align}
&H^{\mu\nu}_{\lambda'_N,\lambda_N}=\nonumber\\
&-i\int {\rm d}^4x \,e^{-i(q\cdot x)}\langle N(p',\lambda'_N)|T[J^\mu(x)J^\nu(0)|N(p,\lambda_N)\rangle,
\label{DVCS-tensor}
\end{align}
where the four-vector index $\mu$ $(\nu)$ refers to the virtual (real) photon, and
$\lambda_N$ ($\lambda'_N$) is the helicity of the incoming (outgoing) nucleon.
The DVCS amplitude is obtained from the DVCS tensor in Eq.~(\ref{DVCS-tensor}) by contracting it with the photon polarization vectors as
\begin{eqnarray}
T_{\lambda'_\gamma\lambda'_N,\lambda_\gamma\lambda_N}=
\varepsilon_\mu(q,\lambda_\gamma)\varepsilon^*_\nu(q',\lambda'_\gamma)
H^{\mu\nu}_{\lambda'_N,\lambda_N},
\end{eqnarray}
where $\lambda_\gamma$ ($\lambda'_\gamma$) denotes the helicity
 of virtual (real) photons respectively.

The DVCS amplitude for unpolarized nucleon and at leading order in $Q$ can be parametrized  as
\begin{align}
&T_{\lambda'_\gamma\lambda'_N,\lambda_\gamma\lambda_N}=
\varepsilon_\mu(q,\lambda_\gamma)\varepsilon^*_\nu(q',\lambda'_\gamma)\frac{(-g^{\mu\nu}_\perp)}{2}\nonumber\\
&\times\left[
\bar u(p',\lambda'_N) \, \gamma \cdot n \, u(p,\lambda_N)\sum_q e^2_q
C^q\right.\nonumber\\
&\left.\hspace{0.5 truecm}-\bar u(p',\lambda'_N)u(p,\lambda_N)\frac{1}{M_N}\sum_q e^2_q
 F^q\right],
\label{lo-ampl2}
\end{align}
where   we introduced the lightlike vector $n^\mu=1/(\sqrt{2}P^+)(1,0,0,-1)$, with $P=(p+p')/2$, and the symmetric tensor $g^{\mu\nu}_\perp=g^{\mu\nu}-n^\mu\tilde p^\nu-n^\nu\tilde p^\mu$, with $\tilde p^\mu=P^+/\sqrt{2}(1,0,0,1)$. Furthermore, the light-front component for a generic four-vector $a^\mu$ is defined as $(a^0+a^3)/\sqrt{2}$.
In Eq.~(\ref{lo-ampl2}), the invariant amplitudes $C^q$ and $ F^q$ are given by
\begin{align}
C^q(\xi,t)=&\int_{0}^1 {\rm d}x\left[H^{(+)}(x,\xi,t)+E^{(+)}(x,\xi,t)\right]\nonumber\\
&\times\left[\frac{1}{x-\xi+i\epsilon}+\frac{1}{x+\xi-i\epsilon}\right]\nonumber\\
&=\int_{-1}^1 {\rm d}x\frac{H^{(+)}(x,\xi,t)+E^{(+)}(x,\xi,t)}{x-\xi+i\epsilon},\label{eq:Campl}\\
F^q(\xi,t)=&\int_{0}^1 {\rm d}xE^{(+)}(x,\xi,t)\left[\frac{1}{x-\xi+i\epsilon}+\frac{1}{x+\xi-i\epsilon}\right]\nonumber\\
&=\int_{-1}^1 {\rm d}x\frac{E^{(+)}(x,\xi,t)}{x-\xi+i\epsilon},\
\label{eq:Fampl}
\end{align}
with the skewedness variable defined as $\xi=Q^2/(2s+Q^2)$.
$H^{(+)}(x,\xi,t)=H^q(x,\xi,t)-H^q(-x,\xi,t)$ 
denotes the singlet ($C=+1$) combination of nucleon helicity-conserving GPDs, 
and analogously for the nucleon helicity-flip GPD $E^{(+)}$.
The invariant amplitudes and the GPDs in Eqs.\eqref{eq:Campl} and \eqref{eq:Fampl} depend also on the renormalization scale $\mu^2$ which is not explicitly displayed and it is identified with the hard scale 
of the process $Q^2$.
\noindent
In the following we will consider the  invariant amplitude $ F^q$ in the $\nu-t$ plane at fixed $Q^2$, with $\nu=(s-u)/4M_N=Q^2/4M_N\xi$.
In this plane, $ F^q$ satisfies the following fixed-$t$ subtracted relation~\cite{Anikin:2007yh,Diehl:2007jb}
\begin{align}
 F^q(\nu,t)= F^q(0,t)+\frac{\nu^2}{\pi}\int_{\nu_0}^\infty\frac{{\rm d}\nu'^2}{\nu'^2}\frac{\mbox{Im}  F^q(\nu',t)}{\nu'^2-\nu^2},
\label{DRnu}
\end{align}
where  the lower limit of integration is $\nu_0=Q^2/4M_N$ and 
the nucleon pole term residing in this point may be 
considered separately.
Following   Refs.~\cite{Anikin:2007yh,Radyushkin:2011dh}, we can relate the subtraction function $ F^q(0,t)$  to the D-term form factor $D^q(t)$~\cite{Polyakov:1999gs} as follows
\begin{align}
\label{eq:subtrfunc}
F^q(0,t)=2\int_{-1}^{+1}{\rm d}z\frac{D^q(z,t)}{1-z}=4D^q(t).
\end{align}
The  dispersive representation for the D-term form factor  $D^q(t)$ of Eq.~\eqref{eq:subtrfunc} is obtained by applying  unsubtracted DRs, this time in the
variable $t$:
\begin{align}
 F^q(0,t)=&\frac{1}{\pi}\int_{4m_\pi^2}^{+\infty}{\rm d}t'
\frac{\mbox{Im}_t  F^q(0,t')}{t'-t}\nonumber\\
+&\frac{1}{\pi}\int_{-\infty}^{-a}{\rm d}t'
\frac{\mbox{Im}_t  F^q(0,t')}{t'-t}
\label{DRt}.
\end{align}
The imaginary part in the integral from $4m_\pi^2\rightarrow+\infty$ in Eq.~(\ref{DRt}) is saturated by the possible intermediate states for the $t$-channel process, which lead to cuts along the positive-$t$ axis. 
For low values of $t$, the $t$-channel discontinuity is dominated by $\pi\pi$ intermediate states.
The second integral in Eq.~(\ref{DRt}) extends from $-\infty$ to $-a=-2(m_\pi^2+2M_Nm_\pi)-Q^2$.
As we are interested in evaluating Eq.~(\ref{DRt}) for large $Q^2$ values and small (negative) values of $t$ ($|t|\ll a$), the integral from $-\infty\rightarrow -a$ is suppressed, and will be neglected in this work.
Consequently, we shall saturate the integral in Eq.~(\ref{DRt}) by the contribution of $\pi\pi$ intermediate states, which turns out to be a good approximation for small $t$.

\noindent
Using the expansion of the D-term $D(z,t)$  in  terms of Gegenbauer
polynomials $C^\nu_k$ for $\nu=3/2$, the solutions of the leading-order ERBL evolution equations, one obtains the following series for the D-term form factor 
\begin{equation}
D^q(t)=\sum_{\scriptstyle n=1 \atop \scriptstyle {\rm odd}}^{\infty}d^q_n(t).
\label{seriesD}
\end{equation}
In the following, we will  explicitly evaluate the contribution from the $n=1$ term in~\eqref{seriesD}.
\section{$t$-channel dispersion relations for the D-term form factor}
\label{sect:3}
The  invariant amplitudes $ F^q(\nu,t)$ and $C^q(\nu,t)$ are related to the $t$-channel helicity amplitude by~\cite{Diehl:1998dk,Diehl:2002yh}
\begin{eqnarray}
&&T^t_{\lambda_{\bar N}\lambda_N,\lambda_\gamma\lambda'_\gamma}=
\varepsilon_\mu(q_t,\lambda_\gamma)\varepsilon_\nu(q'_t,\lambda'_\gamma)
T^{t\,\mu\nu}_{\lambda_{\bar N}\,\lambda_{N}}
\nonumber\\
&&=
\varepsilon_\mu(q_t,\lambda_\gamma)\varepsilon_\nu(q'_t,\lambda'_\gamma)
\frac{(-g^{\mu\nu}_\perp)}{2}\nonumber\\
&&\times\left[\bar u(p_t,\lambda_N) \gamma^+v(p'_t,\lambda_{\bar N})\frac{1}{\tilde{\Delta}^+} \sum_q e^2_q
C^{q}\right.\nonumber\\
&&\hspace{0.5 truecm}\left. -\bar u(p_t,\lambda_N)  v(p'_t,\lambda_{\bar N})\frac{1}{M_N}\sum_q e^2_q
F^{ q}\right],
\label{tch-a}
\end{eqnarray}
where $\tilde{\Delta}^+=\dfrac{{p'}_{t}^{+} -p_{t}^{+}}{2}$, and the hadronic tensor $T^{t\,\mu\nu}_{\lambda_{\bar N}\,\lambda_{N}}$ is defined as
\begin{align}
&T^{t\,\mu\nu}_{\lambda_{\bar N}\,\lambda_{N}}\nonumber\\
&\hspace{-0.2 truecm}=-i\int {\rm d}^4x e^{-i(q\cdot x)}
\langle N(p_t, \lambda_N),\, \bar N(p'_t,\lambda_{\bar N})|T[J^\mu(x)J^\nu(0)|0\rangle.
\label{tch-tensor}
\end{align}
In the c.m. system of the $t$-channel process $\gamma^*\gamma\rightarrow N\bar N$
we choose the real photon momentum $q'_t$ (helicity $\lambda'_\gamma$) 
to point in the $z$ direction and the nucleon momentum  $p_t$ in the $xz$ plane at an angle $\theta_t$ 
with respect to the $z$ axis, i.e. $p_t^\mu=(E,p_t\sin\theta_t,0,p_t\cos\theta_t)$ with $p_t=|\vec p_t|=\sqrt{t/4-M_N^2}$.
In this framework, the $t$-channel helicity amplitude in Eq.~(\ref{tch-a}) can be written as
\begin{align}
T^t_{\lambda_{\bar N}\lambda_{N},\,\lambda_\gamma\lambda'_\gamma}=&\delta_{\lambda_\gamma\lambda'_\gamma}
\delta_{\lambda_N\lambda_{\bar N}}\left[(-1)^{1/2+\lambda_N}\frac{M_N}{\sqrt{2}\tilde\Delta^+}\cos\theta_t\sum_qe^2_qC^q\right.\nonumber\\
&\left.\hspace{1.2 truecm}+(-1)^{1/2+\lambda_N}\sqrt{\frac{t}{4M_N^2}-1}\sum_qe^2_q F^q\right]\nonumber\\
&+\delta_{\lambda_\gamma\lambda'_\gamma}
\delta_{-\lambda_N\lambda_{\bar N}}
\frac{\sqrt{t}}{2\sqrt{2}\tilde\Delta^+}\sin\theta_t\sum_qe^2_qC^q.
\label{hel-a}
\end{align}
Since the dispersion integral in Eq.~(\ref{DRt}) runs along the line $\nu=0$,
we are interested to $\mbox{Im}_t  F^q(0,t)$ in Eq.~(\ref{hel-a}).
The relation between
the scattering angle in the $t$-channel and the invariant $\nu$ and $t$ is
$\cos\theta_t= 4 M_N \nu/[\beta_N (t+Q^2)]$ with $\beta_N =\sqrt{1-4M_N^2/t}$.
Therefore $\nu=0$ corresponds to $90^o$ scattering for the $t$-channel process. In this limit,
the relations~\eqref{hel-a} reduce to
\begin{align}
&T^t_{1/21/2,\,11}(t,\theta_t=90^o)=-\sqrt{\frac{t}{4M_N^2}-1}\sum_qe^2_q F^q(0,t),
\label{hel-b}
\\
&T^t_{1/2-1/2,\,11}(t,\theta_t=90^o)=\frac{\sqrt{t}}{2\sqrt{2}\tilde\Delta^+}\sum_qe^2_qC^q(0,t).
\label{hel-c}
\end{align}
The imaginary part of the $t$-channel Compton amplitude is determined by using unitarity relation, and taking into account the dominant contribution coming from $\pi\pi$ intermediate states.
Following the derivation in App. B of Ref.~\cite{Drechsel:1999rf}, we start by decomposing the $t$-channel helicity amplitude for $\gamma^*\gamma\rightarrow\bar N N$ into a partial wave series,
\begin{equation}
T_{\lambda_{\bar N} \lambda_{N}, \, \lambda_{\gamma} \lambda'_\gamma}^t 
(\nu, t) \;=\; \sum_J {{2 J + 1} \over {2}} \;
T_{\lambda_N \lambda_{\bar N}, \, \lambda'_{\gamma}
  \lambda_\gamma}^{J(\gamma^*\gamma\rightarrow N\bar N)}
(t)\; d^J_{\Lambda_N \Lambda_\gamma} (\theta_t) \;,
\label{eq:pwgagannbar}
\end{equation}  
where $\Lambda_\gamma=\lambda'_\gamma-\lambda_\gamma$,
 $\Lambda_N=\lambda_N-\lambda_{\bar N}$, and $d^J_{\Lambda_N \Lambda_\gamma}$ are Wigner $d$-functions.
The unitarity relation reads 
\begin{align}
&2\,\mathrm{Im} T^{\gamma^* \gamma \, \rightarrow \, N \bar N} =\nonumber\\
&
{1 \over { {(4 \pi)}^2 } } 
\frac{p_{\pi}}{\sqrt{t}} 
\int {d \Omega}_{{\pi}}\,
\left [ \, T^{\gamma^* \gamma \, \rightarrow \, \pi \pi}\, \right ] 
\cdot \left [ \, T^{\pi \pi \, \rightarrow \, N \bar N}\, \right ] ^ { \ast },
\end{align}
where $p_\pi=|\vec p_\pi|= \sqrt{t/4 - m_\pi^2}$ is the  c.m. momentum of the pion.
The partial wave expansion for $\gamma^* \gamma \rightarrow \pi
\pi$ reads
\begin{align}
T^{\gamma^* \gamma \, \rightarrow \, \pi \pi}
_{{\Lambda}_{\gamma}} (t, \theta_{\pi \pi}) =&
\sum_{\scriptstyle J=0 \atop \scriptstyle {\rm even}} \, {{2J+1} \over 2} 
T^{J \;(\gamma^* \gamma \, \rightarrow \, \pi \pi)}
_{{\Lambda}_{\gamma}}(t) \nonumber\\
&\times
\sqrt {(J-\Lambda_{\gamma})! \over {(J+\Lambda_{\gamma})!}} \, \cdot \,
P_J^{\Lambda_{\gamma}}(\cos \theta_{\pi \pi}).
\label{eq:partialgagapipi}
\end{align}
Furthermore, the partial wave expansion for $\pi \pi \rightarrow N \bar N$ reads
\begin{align}
T^{\pi \pi \, \rightarrow \, N \bar N}
_{\Lambda_N} (t, \Theta) =&
\sum_{J} \, {{2J+1} \over 2} \; 
T^{J \; (\pi \pi \, \rightarrow \, N \bar N)}
_{\Lambda_N}(t) \nonumber\\
&
\times\sqrt {(J-\Lambda_N)! \over {(J+\Lambda_N)!}} \, \cdot \,
P_J^{\Lambda_N}(\cos \Theta) \;.
\label{eq:partialpipinnbar}
\end{align}
Combining Eqs.~(\ref{eq:partialgagapipi}) and (\ref{eq:partialpipinnbar}),
we can now construct the imaginary parts of the
Compton $t$-channel partial waves, 
\begin{align}
&2\,\mathrm{Im} T^{J \;(\gamma^* \gamma \, \rightarrow \, N \bar N)}
_{\lambda_{\bar N} \lambda_{N}, \, \lambda_{\gamma} \lambda'_\gamma}(t)\nonumber\\
&={1 \over {(8 \pi)} } 
{p_{\pi} \over \sqrt{t}}
\left [ \, T^{J \;(\gamma^* \gamma \, \rightarrow \, \pi \pi)}
_{\Lambda_\gamma}(t)\, \right ] 
\left [ \, T^{J \;(\pi \pi \, \rightarrow \, N \bar N)}
_{\Lambda_N }(t)\, \right ] ^ { \ast }.
\label{eq:partialtunit}
\end{align}
For the calculation of ${\rm Im} F^q(0,t)$ from Eq.~(\ref{hel-b}), 
we should consider Eq.~(\ref{eq:partialtunit}) for $\Lambda_\gamma=0$ and $\Lambda_N=0$.
\newline
\noindent
The partial wave amplitudes $T^{J \;(\pi \pi \, \rightarrow \, N
  \bar N)} _{\Lambda_N=0}$ of
Eq.~(\ref{eq:partialpipinnbar}) are related to the
amplitudes $ f^J_{+}(t)$ of Frazer and Fulco
\cite{Frazer:1960zza} by the relation
\begin{align}
T^{J \;(\pi \pi \, \rightarrow \, N \bar N)}_{\Lambda_N = 0}(t)\;
= \frac{16 \pi}{p_t}\,(p_t \; p_\pi)^J \,
f^J_+ (t)\ .\nonumber
\label{eq:frazerfulco}
\end{align}
\newline
\noindent
The reaction $\gamma^* \gamma \, \rightarrow \, \pi \pi$ at large $Q^2$ and small $t$ can be described in a factorized form~\cite{Diehl:1998dk,Diehl:2000uv}, as the convolution of a short-distance contribution, $\gamma^*\gamma\rightarrow q \bar q$, 
perturbatively
calculable, and nonperturbative matrix elements describing the exclusive fragmentation of a $q\bar q$ pair into two-pion. These nonperturbative functions correspond to two-pion generalized distribution amplitudes (GDAs), denoted as $\Phi_q^{\pi\pi}$.
For transversely polarized photons,  the helicity amplitude for $\gamma^*\gamma\rightarrow \pi\pi$ at leading twist reads~\cite{Diehl:1998dk}
\begin{equation}
T^{\gamma^*\gamma\rightarrow\pi\pi}_{\Lambda_\gamma=0}=
\frac{1}{2}\sum_qe^2_q\int_0^1{\rm d}z\frac{2z-1}{z(1-z)}\Phi_q^{\pi\pi}(z,\zeta,t),
\label{eq:ggpp}
\end{equation}
where $z$ is
the fraction of light-cone momentum carried by the quark 
with respect to the 
pion pair and  $\zeta$ is the fraction of light-cone momentum carried by the pion with respect to the pion pair, i.e.
\begin{align}
\zeta=\frac{1+\beta\cos\theta_{\pi\pi}}{2},\qquad \beta=\sqrt{1-\frac{4m_\pi^2}{t}}.
\end{align}
\newline
\noindent
In Eq.~(\ref{eq:ggpp}), we can distinguish the neutral and charged pion channel production.
The process $\gamma^*\gamma\rightarrow \pi^+\pi^-$ is only sensitive to the $C$ even part of  $\Phi_q^{\pi^+\pi^-}$, since the initial two-photon state has positive $C$-parity. On the other side,  the $\pi^0\pi^0$ pair has positive $C$-parity as well, so that $\Phi_q^{\pi^0\pi^0}$ has no $C$-odd part at all. 
Isospin invariance implies that
the pion pair is in a state of zero isospin and 
$\Phi_u^+=\Phi_d^+$, where the index $+$ denotes the $C$-even contribution.
As a result, we have
\begin{equation}
\Phi_q^{\pi^+\pi^-}=\Phi_q^{\pi^0\pi^0}=\Phi_q^+.
\end{equation}

The GDAs have the following  partial wave expansion~\cite{Diehl:2000uv,Polyakov:1998ze,Kivel:1999sd}
\begin{equation}
\label{eq:series1}\Phi_q^+ = 6 \, z(1-z)
\sum_{\scriptstyle n=1 \atop \scriptstyle {\rm odd}}^{\infty}
\sum_{\scriptstyle l=0 \atop \scriptstyle {\rm even}}^{n+1}
B_{nl}^q(t)\, C_n^{(3/2)}(2 z-1)\, P_l(2\zeta-1),
\end{equation}
where $C_n^{(3/2)}$ are Gegenbauer polynomials and the expansion coefficients $B^q_{nl}$ contains a dependence on the factorization scale, which is not shown explicitly. 
The expansion of the $\zeta$-dependence in Legendre  polynomials is directly related to the partial-wave expansion of the two-pion system.
As a matter of fact, one can rewrite the polynomials $P_l(2\zeta-1)=P_l(\beta\cos\theta_{\pi\pi})$ in terms of $P_k(\cos\theta_{\pi\pi})$ with $k\le l$, with  the series (\ref{eq:series1}) transforming in
\begin{equation}
\label{eq:series2}
\Phi_q^+ = 6 \, z(1-z)
\sum_{\scriptstyle n=1 \atop \scriptstyle {\rm odd}}^{\infty}
\sum_{\scriptstyle l=0 \atop \scriptstyle {\rm even}}^{n+1}
\tilde{B}_{nl}^q(t)\, C_n^{(3/2)}(2 z-1)\, P_l(\cos\theta_{\pi\pi}),
\end{equation}
where the coefficients $\tilde B_{nl}$ are linear combinations of the form
\begin{equation}
\tilde B_{nl}=\beta^l[B_{nl}+c_{l,l+2}B_{n,l+2}+\cdots +c_{l,n+1}B_{n,n+1}],
\end{equation}
with polynomials $c_{l,l'}$ in $\beta^2$.

\noindent
Inserting Eqs.~(\ref{eq:ggpp}) and (\ref{eq:series2}) in the partial wave expansion of the helicity amplitude in Eq.~(\ref{eq:partialgagapipi}), one finds
\begin{align}
T^{J \;(\gamma^* \gamma \, \rightarrow \, \pi \pi)}
_{{\Lambda}_{\gamma}=0}(t)=\sum_q \,e^2_q\,
T^{J \;(\gamma^* \gamma \, \rightarrow \, q \bar q)}
_{{\Lambda}_{\gamma}=0}(t)
\end{align}
with
\begin{align}
&T^{J \;(\gamma^* \gamma \, \rightarrow \, q \bar q)}
_{{\Lambda}_{\gamma}=0}(t)=
\frac{6}{2J+1}\nonumber\\
&\times\sum_{\scriptstyle  n={\rm  max}(1,J-1)\atop  \scriptstyle {\rm odd}}^\infty
\int_0^1{\rm d}z\,(2z-1)\tilde B^q_{nJ}(t) C_n^{(3/2)}(2z-1).
\label{pipi-series}
\end{align}
Inserting the partial wave expansion of Eq.~(\ref{eq:partialtunit}) into Eq.~(\ref{hel-b}), 
we can finally express the $2\pi$ $t$-channel contribution to
${\rm Im}_t  F^q(\nu=0,t)$ by the partial wave amplitudes for the reactions $\gamma^*\gamma\rightarrow\pi\pi$ and $\pi\pi\rightarrow N\bar N$
\begin{align}
&{\rm Im}_t F^{q\,(\pi\pi)}
=-\frac{M_Np_\pi}{\sqrt{t}\,p_t^2}\nonumber\\
&\times
\sum_{\scriptstyle J \atop \scriptstyle {\rm even}} \, {{2J+1} \over 2} \; 
(-1)^{J/2}\frac{(J-1)!!}{J!!}(p_tp_\pi)^J\,T^{J \;(\gamma^* \gamma \, \rightarrow \, q \bar q)}_{\Lambda_\gamma=0}
f^{J*}_+(t).
\label{eq:finalt}
\nonumber\\
\end{align}
For the numerical estimate, we restrict ourselves to the $S$- and $D$-wave  contributions in Eq.~(\ref{eq:finalt}).
The partial-wave amplitudes of  the $\pi \pi \rightarrow N \bar N$ subprocess are taken
from the work of H\"ohler and collaborators~\cite{hoehler:1983}, 
in which the lowest $\pi\pi \rightarrow N \bar N$ partial wave 
amplitudes were constructed from a partial wave solution of
pion-nucleon scattering, by use of the $\pi \pi$ phaseshifts of
Ref.~\cite{Froggatt:1977hu}. In Ref.~\cite{hoehler:1983}, the 
$\pi \pi \rightarrow N \bar N$ amplitudes are given 
for $t$ values up to $t \approx 40 \cdot m_\pi^2 \approx$
0.78~GeV$^2$, which is taken as upper limit of integration in the $t$-channel dispersion integral~\eqref{DRt}.
The latter value corresponds to the onset of inelasticities in the
$\pi\pi$
phase shifts.
\newline
\noindent
The $S$- and $D$-wave amplitudes of  the $\gamma^*\gamma\rightarrow \pi \pi $ subprocess
are calculated from Eq.~\eqref{pipi-series}, taking into account only the $n=1$ term.
This corresponds to restrict our dispersion evaluation to the $d^q_1(t)$ term in the series~\eqref{seriesD}.
The  two-pion GDAs are calculated through dispersion relations using the Omn\`es representation which was first discussed in Ref.~\cite{Polyakov:1998ze} and further used in Refs.~\cite{Kivel:1999sd,LehmannDronke:1999aq,LehmannDronke:2000xq,Warkentin:2007su}.
Following the derivation of Ref.~\cite{Warkentin:2007su}, the results for the $S$- and $D$-wave coefficients reads
\begin{eqnarray}
\tilde B^q_{10}(t)&=&-B^q_{12}(0)\frac{3C-\beta^2}{2}f_0(t)\label{eq:b10}\\
\tilde B^q_{12}(t)&=&\beta^2B^q_{12}(0)f_2(t),\label{eq:b12}
\end{eqnarray}
where the Omn\`es functions $f_{0,2}$ can be related to $\pi\pi$ phase-shifts $\delta_{0,2}^0(t)$ using Watson theorem and dispersion relations derived in~\cite{Polyakov:1998ze}:
\begin{align}
f_l(t)=\exp\left[\frac{t}{\pi}\int_{4m_\pi^2}^\infty{\rm d}t'\frac{\delta_l^0(t')}{t'(t'-t-i\epsilon)}\right].\label{fpipi}
\end{align}
In Eq. (\ref{eq:b10}), the constant $C$ is taken from Refs.~\cite{Kivel:1999sd}, using the estimate from the instanton model~\cite{Polyakov:1998td} at low energies,
$C=1+bm_\pi^2+{\cal O}(m_\pi^4)$ with $b\approx-1.7$ GeV$^{-2}$, while the coefficient $B_{12}(0)$ is obtained using the crossing relations between the quark 2$\pi$DA's and the corresponding parton distributions in the pion, i.e.
\begin{align}
\label{b12-coeff}
B^q_{12}(0)=\frac{10}{9}\int{\rm d}x\, x\frac{1}{N_f}\sum_f[q_\pi^f(x)+\bar q_\pi^f(x)].
\end{align}
As final result, taking into account only the contribution with  $J=0$ and $J=2$, Eq.~\eqref{eq:finalt} simplifies to
\begin{align}
&{\rm Im}_t F^{q\,(\pi\pi)}
=\frac{3M_Np_\pi}{2\sqrt{t}\,p_t^2}B^q_{12}(0)\nonumber\\
&\times
\left[(3C-\beta^2)f_0(t)\,f^{0*}_+(t)+(p_\pi p_t)^2\beta^2 f_{2}(t)f^{2*}_+(t)\right].
\label{eq:finalt2}
\end{align}
In Eq.~\eqref{eq:finalt2}, the dependence on the renormalization scale enters only through the coefficient $B^q_{12}$ evaluated at $t=0$, and therefore is factorized from the $t$ dependence of the amplitude. Furthermore, the coefficients $B_{12}^q$ evolve in the same way as the quark momentum fraction in the pion, in accordance with Eq.~\eqref{b12-coeff}.

\section{Results}\label{sec:4}
    
\begin{figure}[t]
\begin{center}
\epsfig{file=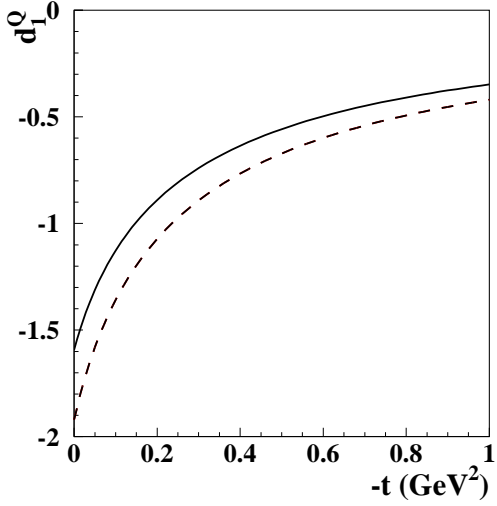, width=\columnwidth}
\end{center}
\caption{$d_1^Q$ as function of $t$, obtained with different inputs for the quark distributions in the pion $q_\pi^f$. Solid curve: results with $q_\pi^f$ from Ref.~\cite{Owens:1984zj}. 
Dashed  curve: results with $q_\pi^f$ from Ref.~\cite{Gluck:1991ey}. The results refer to the scale $Q^2=4$ GeV$^2$.}
\label{fig:2}
\end{figure}

In Fig.~\ref{fig:2} we present the dispersive predictions for $d_{1}^Q=\sum_q d_{1}^{q}(t)$ as function of $t$, with the sum over  flavors restricted to up and down quarks.
The  solid  and dashed curve are obtained using as input in Eq.~\eqref{b12-coeff}  the parametrization of the pion distributions at $Q^2=4$ GeV$^2$ from Ref.~\cite{Owens:1984zj} and~\cite{Gluck:1991ey}, respectively.
The different inputs for the pion distributions change the results by an overall normalization factor, without affecting the $t$ dependence.
As outlined above, the $Q^2$ dependence enters only  through the quark momentum fraction of the pion, which changes   only by a few percent in the range of $Q^2=[1,10]$ GeV$^2$.
At $t=0$, we find $d_1^Q=-1.59$ and $d^Q_1(0)=-1.92$ for the solid and dashed curve in Fig.~\ref{fig:2}, respectively.
These values compare with the results obtained, at a low normalization scale, in the $\chi$QSM~\cite{Goeke:2007fp}, $d_{1}^Q(0)=-2.35$, in the Skyrme model~\cite{Cebulla:2007ei}, $d_1^Q(0)= -4.48$, and in a recent calculation 
with effective light-front wave functions from a Regge-improved diquark model~\cite{Mueller:2011bk},
$d_1^q(0)=-2.01$.
 \newline
 \noindent
 Among the form factor in Eq.~\eqref{eq:dterm},  $d_1^Q(t)$ aroused a particular interest, as it  enters in the parametrization of the quark part of the energy momentum tensor of QCD,
 and provides information on how strong forces are distributed and stabilized in the nucleon~\cite{Polyakov:2002yz}.
In all theoretical studies so far as well as in the present dispersive calculation,  $d^{Q}_1(t)$
at zero-momentum transfer $t=0$
 is found to have a negative sign. The negative values of this constant has a deep relation to
the spontaneous breaking of the chiral symmetry in QCD~\cite{Polyakov:2002yz,Polyakov:2002wz}, and has also an appealing connection with the criterion of stability of the nucleon~\cite{Goeke:2007fp}. \begin{figure}[t]
\begin{center}
\epsfig{file=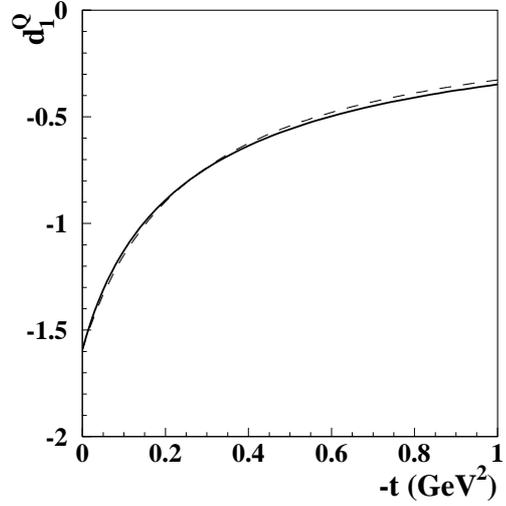, width=\columnwidth}
\end{center}
\caption{Model calculation of $d_1^Q(t)$ (solid curve) in comparison with the function in Eq.~\eqref{eq:fit}.}
\label{fig:4}
\end{figure}
\newline
\noindent
Furthermore, $d_1$ determines the behavior of the D-term form factor in the asymptotic limit $\mu^2\rightarrow \infty$.
In this limit,
all the terms with $n>1$
in the series~\eqref{seriesD} go to zero, and one has
\begin{align}
D^{Q,\, {\rm as}}(t)= d(t)\frac{3N_f}{3N_f+16},
\end{align}
where $d(t)=d_1^Q(t)+d_1^G(t)$ is the total, scale-independent, contribution from quark and gluon.

In the dispersive calculation, the asymptotic limit of $D^Q(t)$ can be obtained from the asymptotic limit of $B_{12}(0)$ in Eq.~\eqref{b12-coeff}, i.e.
\begin{align}
\label{b12-coeff-as}
B_{12}^{Q,\, {\rm as}}(0)=\frac{10}{9}\frac{3N_f}{3N_f+16}.
\end{align}
As a result, $d(t)$ has the same $t$-dependence of $d_{1}^Q(t)$ shown in Fig.~\ref{fig:2}, and differs only for the value at $t=0$ which is found $d(0)=-3.32$.

In most of phenomenological studies of DVCS, the $t$ dependence of D-term form factor is parametrized by a dipole function~\cite{Guidal:2013rya}.
However, the dispersive results  favor a different functional form, as shown
in Fig.~\ref{fig:4}  where we compare the result for $d_1^Q$ as function of $t$ with the following parametrization 
\begin{align}
F_D=\frac{d_{1}^Q(0)}{[1- t/(\alpha M_D^2)]^\alpha},
\label{eq:fit}
\end{align}
with $M_D=0.487 $ GeV and $\alpha=0.841$.
\begin{figure}[t]
\begin{center}
\epsfig{file=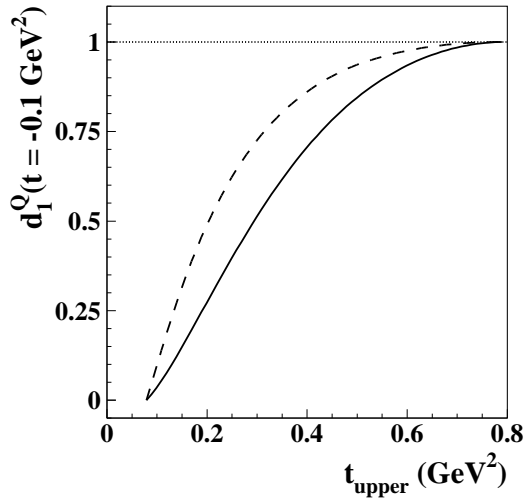, width=\columnwidth}
\end{center}
\caption{The results at $t=-0.1$ GeV$^2$ for the unsubtracted (solid curve)
and the subtracted (dashed curve)
$t$-channel dispersion integrals in Eq.~\eqref{DRt} and \eqref{DRtsub}, respectively, are shown as function of the upper integration limit $t_{{\rm upper}}$. Both results are normalized
to their respective values at
$t_{{\rm upper}}=0.78$ GeV$^2$. }
\label{fig:5}
\end{figure}

In Fig.~\ref{fig:5} we show the convergence of the
$t$-channel integral from $4 m_\pi^2$ to $\infty$
in the unsubtracted DR of Eq.~\eqref{DRt} for $t=-0.1$ GeV$^2$. We
do so by calculating the dispersion integral as function of the
upper integration limit $t_{{\rm upper}}$ and by showing the ratio to the
integral for
$t_{{\rm upper}}$= 0.78 GeV$^2$. The latter value corresponds
to the the onset of inelasticities in the
$\pi\pi$
phase shifts.
 One  sees from Fig.~\ref{fig:5} that the
unsubtracted
$t$-channel DR shows only a slow convergence.

In order to improve the convergence of the dispersion integral, we may introduce subtracted DRs, with the subtraction constant  at $t=0$:
\begin{align}
 D^q(0,t)=&D^q(0)+\frac{t}{4\pi}\int_{4m_\pi^2}^{+\infty}{\rm d}t'
\frac{\mbox{Im}_t  F^q(0,t')}{t'(t'-t)},\label{DRtsub}
\end{align}
where we omitted the contribution from the negative $t$-channel cut.
In Fig.~\ref{fig:5} we see that the  the subtracted dispersion integral converges faster, reaching its final value around $t\approx 0.6$ GeV$^2$.
The price to pay is the appearance in Eq.~\eqref{DRtsub} of the subtraction constant that has to be fitted to experimental data.
To have a rough indication of the contribution expected  above the inelastic threshold, we extended the integration up to $t_{{\rm upper}}=1.78$ GeV$^2$, including the inelasticities in the $\pi\pi$ phase shifts and approximating the $\pi N$ partial-wave amplitudes with the Born contribution. The  results of  the unsubtracted DRs are affected by $\sim10 \%$, while the subtracted dispersion integrals are quite stable and change just by a few percent. 

\section{Conclusions}
We have presented a dispersive representation for the quark contribution to the D-term form factor in hard exclusive reactions in terms of unsubtracted $t$-channel dispersion relations.
The unitarity relation for the $t$-channel amplitudes is saturated with two-pion intermediate states, taking
into account the contribution from $S$-and $D$-wave intermediate states in the numerical estimate.
The input for the imaginary part of the dispersion relation are the two-pion GDAs, determined through  the first-$x$ moment of the flavor-singlet pion PDFs,
the  $\pi\pi$ phase shifts up to the inelastic threshold, and the partial waves for the $\pi\pi\rightarrow N\bar N$ amplitudes obtained from dispersion theory  by analytical continuation of $\pi N$ scattering.
 We found that the $t$ and $Q^2$ dependence of the D-term form factor  are disjoined.
The $t$-dependence is not trivial and it does not follow a dipole behavior as normally assumed in phenomenological parametrizations.
 On the other hand, the $Q^2$ dependence  enters only in 
the normalization point at $t=0$, which is proportional to the first $x$-moment of the flavor-singlet pion PDFs.
The value at $t=0$ is also compatible with estimates in chiral-quark soliton model anda Regge-improved diquark model.
In order to improve the convergence of the dispersion integral, we also discussed subtracted dispersion relations, which can be used to determine the $t$-dependence of the D-term form factor, 
but leave  the value at $t=0$ as free parameter to be fitted to experimental data.

\section*{Acknowledgements}
{The authors are  thankful to P. Schweitzer for a careful reading of the manuscript and instructive discussions.  B.P. is also grateful to D. M\"uller, H. Moutarde and O. Teryaev for stimulating comments.
This work was supported in part by the European Community Joint Research Activity ``Study of Strongly Interacting Matter'' (acronym HadronPhysics3, Grant Agreement n. 283286)
under the Seventh Framework Programme of the European Community.


\end{document}